\documentstyle[aps,preprint]{revtex}
\begin{document}
\draft

\title{Soluble Infinite-Range Model of Kinetic Roughening}
\author{M. Marsili\cite{addr} and A. J. Bray}
\address{Department of Physics and Astronomy, The University, 
Manchester M13 9PL, UK}
\date{\today}
\maketitle

\begin{abstract}
A modified Kardar-Parisi-Zhang (KPZ) equation is introduced, and solved 
exactly in the infinite-range limit. In the low-noise limit the system 
exhibits a weak-to-strong coupling transition, rounded for non-zero noise, 
as a function of the KPZ non-linearity. The strong-coupling regime is 
characterised by a double-peaked height distribution in the stationary 
state. The nonstationary dynamics is quite different from that of the 
stationary state. 
\end{abstract}

\narrowtext


The field of kinetic roughening of surfaces has seen an explosion of 
interest in recent years (see, e.g, \cite{kpzrev,krug}). 
Numerous theoretical models have been proposed, of which the simplest 
non-trivial example is the Kardar-Parisi-Zhang (KPZ) equation \cite{okpz}
\begin{equation}
\partial_t h=\nu \nabla ^2h+\lambda \left( {\vec \nabla h} \right)^2+\eta 
\label{kpz}
\end{equation}
where $h(\vec x,t)$ specifies the height of the surface at point $\vec x$ at 
time $t$, and $\eta(\vec{x},t)$ is a zero mean Gaussian noise with
correlator
$\left\langle {\eta (\vec x,t)\eta (\vec x,t)} \right\rangle=
2D_0\delta ^d(\vec x-\vec x)\delta (t-t)$.
This same equation is believed to describe the statistics of directed 
polymers in a random medium \cite{kpz,kpzrev}, where $h$ has the meaning
of a  free energy.

For spatial dimensionality $d>2$, the KPZ equation is known to exhibit, 
for fixed noise strength $D_0$, a phase transition at $\lambda=\lambda_c$ 
from a `weak-coupling' or `smooth' phase for $\lambda<\lambda_c$ to a 
`strong-coupling' or `rough' phase for $\lambda>\lambda_c$. For $d \leq 2$, 
only the strong-coupling phase exists for any $\lambda>0$. 

Despite intense effort in recent years, the properties of the strong-coupling 
phase are rather poorly understood. In the present Letter, we attempt to  
rectify this by presenting an exact solution of the model in the infinite 
range (or infinite $d$) limit. It turns out, however, that (\ref{kpz}) has 
no stationary state in this limit, exhibiting instead a finite-time 
singularity. By adding an extra, physically-motivated term 
[see (\ref{kpzRSOS}) below], this deficiency is remedied and a stationary 
state obtained.  

The main properties of the model are: (i) In the limit $D \to 0^+$, there 
is a sharp phase transition at a critical coupling $g_c$ from a smooth 
weak-coupling phase to a rough strong-coupling phase; (ii) The strong-coupling 
phase has a two-peaked height distribution in the stationary state, 
with the height difference depending on $g$ but not on $t$; 
(iii) At finite noise strength the transition is rounded, but the 
qualitative features of the weak and strong-coupling regimes are very 
different;   
(iv) The dynamics of the approach {\em to} the stationary state is quite 
different from the dynamics of correlations {\em in} the stationary state, 
contrary to the assumptions of conventional scaling theories.  

We shall adopt an approach which parallels the one used
in mean-field, infinite range, theories for equilibrium statistical mechanics
(e.g.\ in the Sherrington-Kirkpatrick model of spin glasses\cite{SK}), 
introduced in this context originally by Luck and Orland\cite{orland}.
This amounts to assuming, once eq.\ (\ref{kpz}) is discretized
on a lattice, that every site is the nearest neighbor of every other
site. Therefore the number $z$ of nearest neighbors of each site
equals the system size. For Euclidean lattices $z=2d$, so this
approximation is expected to be valid for $d\gg 1$.

Let $h_{i}(t)\equiv h(\vec{x}_i,t)$ where $\vec{x}_i$ is the $i^{\rm th}$
lattice point and let $\sqrt{z}$ be the mesh length of the lattice 
discretization. Let $\langle{h}\rangle_t=\frac{1}{z}\sum_{i=1}^z h_{i}(t)$ be the
average height at time $t$; then 
$\phi_{i}(t)=h_{i}(t)-\langle{h}\rangle_t$ 
is the local height fluctuation on site $i$. 
The discretized form of the Laplacian 
is $\nabla ^2h|_{\vec{x}_i}=-\phi_{i}$, while the nonlinear term
becomes $( {\vec \nabla h} )^2|_{\vec{x}_i}=
(\langle{\phi^2}\rangle_t+\phi_{i}^2)/2$.
After subtracting the equation $\partial_t \langle{h}\rangle_t = 
\lambda\langle{\phi^2}\rangle_t$ for the mean height, (\ref{kpz}) becomes
\begin{equation}
\partial _t\phi =-\nu \phi +
\lambda \left( \phi ^2-\langle{\phi ^2}\rangle_t \right)/2+\zeta 
\label{kpzorl}
\end{equation}
where the site subscript has been suppressed. Here the
noise, in view of our choice of the lattice
discretization, satisfies $\langle{\zeta(t)\zeta(t')}\rangle= 2D\delta(t-t')$ 
with $D=D_0z^{-d/2}$. 

That (\ref{kpzorl}) leads to difficulties at large $t$ can be understood 
by expressing it in the form of a Langevin equation, 
\begin{equation}
\partial _t\phi =-V'(\phi)+\eta\ ,
\label{potential}
\end{equation}
(here and below the prime denotes a derivative)
and noting that the potential $V$ is unbounded from below as 
$\phi\to\infty$. This suggests that there is no stationary state, 
and that a finite-time singularity occurs.
The `self-consistent' time dependence of $V$ which comes 
from the  $\langle{\phi ^2}\rangle_t$ term in (\ref{kpzorl}) make things more 
complex. Numerical integration of the Langevin equation (evolving an 
ensemble of $z\gg 1$ systems simultaneously to compute 
$\langle{\phi ^2}\rangle_t$)
confirm the conclusion of a finite time singularity.
This suggests that the $d \to \infty$ limit of the 
KPZ equation is, as it stands, not well defined.

Since the KPZ equation is usually obtained from a small
gradient expansion, it is reasonable to try to remedy to
the failure of eq.\ (\ref{kpzorl}) by including a further 
term. In order to do this we need to specialize to a particular
model. Indeed while the terms appearing in eq.\ (\ref{kpz})
belong to the small gradient expansion of a large number
of models, which is the origin of the assumed KPZ universality,
the first neglected term differs from model to model. 
For example, in the small gradient expansion for directed
polymers, the next term in (\ref{kpz}) would be $\kappa (\nabla^2 h)^2$.
On the contrary, in models for growing interfaces the next
term is proportional to $({\vec \nabla h})^2\nabla^2 h$ \cite{mmtb}.
Focusing on the latter models, in the following, we will 
discuss the equation
\begin{equation}
\partial _th=\nu \nabla ^2h+\lambda \left( {\vec \nabla h} \right)^2+
\kappa \left( {\vec \nabla h} \right)^2\nabla ^2h+\eta.
\label{kpzRSOS}
\end{equation}
Grouping the first and the third term in the right hand side of 
eq.\ (\ref{kpzRSOS}), this reduces to eq.\ (\ref{kpz}) with a space 
dependent coefficient $\nu$ which is larger
on steep portions of the interface ($\kappa>0$).
This is appropriate for restricted solid-on-solid (RSOS) models 
\cite{kpzrev} for growing interfaces, where, for entropic reasons, 
the fluctuations are suppressed faster 
by the constraint on more inclined surface elements\cite{mmtb}.
The second term also results from the effect
the constraint has on the growth mechanism. Incoming particles are
more likely to stick to the interface, causing it to grow, on flatter
parts of the surface than on inclined ones\cite{mmtb,note}.
Eq.\ (\ref{kpzRSOS}) includes both these effects of the
constraint to the same (second) order in their gradient expansion
\cite{nota}. 

Let us now discuss the infinite range limit
of eq.\ (\ref{kpzRSOS}) for $\kappa>0$. 
Performing the same steps as before leads to 
\begin{equation}
\partial _t\phi=\mu_t-(1+s_t)\phi+g\phi^2-\phi^3+\zeta.
\label{modorl}
\end{equation}
Here we have set $\nu=\kappa=1$ via the transformations 
$t\to t/\nu$ and $\phi\to\phi\sqrt{\nu/\kappa}$. Accordingly
$g=\lambda /(2\sqrt{\nu\kappa})$ and $D=D_0\kappa\nu^{-2} z^{-d/2}$,
where $D_0$ is the strength of the noise in eq.\ (\ref{kpzRSOS}).
Furthermore in this equation we have defined the (time-dependent) 
parameters
\begin{equation}
\mu_t=\langle{\phi^3}\rangle_t-
g\langle{\phi^2}\rangle_t\;\;\;\;\hbox{and}\;\;\;\;
s_t=\langle{\phi^2}\rangle_t
\label{mus}
\end{equation}
which have to be determined self-consistently. The parameter $\mu_t$ 
enforces the condition $\langle{\phi}\rangle_t=0$ for all $t$. The interface 
velocity is $v_t=\partial_t\langle h\rangle_t = gs_t-\mu_t$. 

Eq.\ (\ref{modorl}) can be easily put in the form (\ref{potential})
with a time-dependent potential 
\begin{equation}
V_t(\phi)=-\mu_t\phi+(1+s_t)\frac{\phi^2}{2}-g\frac{\phi^3}{3}
+\frac{\phi^4}{4}
\label{V(phi)}
\end{equation}
which is now bounded from below.

If we assume the parameters $\mu$ and $s$ approach finite limits 
as $t\to\infty$, then the stationary state is given by
\begin{equation}
P_{t=\infty}(\phi)\propto \exp [-V_{\infty}(\phi)/D].
\label{P(phi)}
\end{equation}
On the other hand, the values of $\mu_{\infty}$ and $s_{\infty}$ which 
enter the potential 
$V_{\infty}$ are obtained from eq.\ (\ref{mus}) from the 
distribution $P_{\infty}(\phi)$. 
A stationary state exists if one can find $\mu_{\infty}(g,D)$ and
$s_{\infty}(g,D)$ which solve self-consistently equations (\ref{mus},
\ref{V(phi)}) and (\ref{P(phi)}).

Let us consider first the limit $D \to 0^+$. For small $g$ the potential
$V_{\infty}$ has a single minimum at $\phi^-=0$, 
giving $s_{\infty}=0=\mu_{\infty}$. 
For $g>\sqrt{3}$, 
the potential has a pair of inflection points and for $g>2$ a 
second minimum at $\phi^+=\frac{1}{2}(g+\sqrt{g^2-4})$ appears. 
However, for $g<3/\sqrt{2}$ this minimum has a higher potential than the 
minimum at $\phi=0$, so its occupation is exponentially
suppressed with respect to that of the other minimum for $D \to 0^+$, and  
$s_{\infty}$ remains zero (in fact $s_{\infty}=O(D)$ for $D$ small). 
It is only when, increasing $g$, $V_{\infty}(\phi^+)-
V_{\infty}(\phi^-)$ becomes of 
order $D$ that the weight of the $+$ minimum becomes comparable to
that of the $-$ one. In the limit $D=0^+$ this happens at 
$g_c(0^+)=3/\sqrt{2}=2.12132\ldots$. For $g>g_c$ both minima
contribute to $\langle{\phi^2}\rangle_{\infty}$ so that $s_{\infty}$ 
starts to increase with
$g$ even for $D=0^+$. Since both minima must have non-vanishing 
occupation to ensure $\langle \phi \rangle_{\infty} =0$, $s_{\infty}$ 
and $\mu_{\infty}$ adjust their 
values such that both minima have the same depth. This implies
that $V_{\infty}(\phi)=\bar{V}+\frac{1}{4}(\phi-\phi^-)^2 
(\phi-\phi^+)^2$   where
$\phi^\pm$ are given by ${V'}_{\infty}(\phi^\pm)=0$. Moreover, in this
limit
$P_{\infty}(\phi)=\alpha\delta(\phi-\phi^-)+(1-\alpha)
\delta(\phi-\phi^+)$.
Using $\langle{\phi}\rangle_{\infty}=0$ and evaluating 
$\langle{\phi^2}\rangle_{\infty}$ one 
finds $s_{\infty}=-\phi^-\phi^+$. The solution to these equations yields
$\phi^\pm=\frac{1}{3}g\pm\sqrt{\frac{2}{9}g^2-\frac{1}{2}}$
and 
\begin{equation}
s_{\infty}(g,0^+)=\frac{g^2}{9}-\frac{1}{2},\;\;\;
\mu_{\infty}(g,0^+)=-\frac{g}{3}\left(\frac{g^2}{9}-\frac{1}{2}\right)
\label{D=0+}
\end{equation}
for $g>g_c$. 
The potential at the bottom of both minima is
$\bar{V}=-s^2_{\infty}/4$. These minima are separated by a maximum at
$\phi^o=g/3$ with $V_{\infty}(\phi^o)=\frac{1}{108}g^2(g^2-3)$.

The emerging picture for $D=0^+$ is that of a sharp phase
transition at $g_c$. This separates physically different 
weak-coupling ($g<g_c$) and strong-coupling ($g>g_c$) regimes in 
which the static and dynamic properties differ substantially. 
Note that $D=0^+$ means we are taking the limit $D \to 0$ {\em after} 
the limit $t \to \infty$, in order to establish a stationary state.

Before commenting on the nature of the two phases, it is worth discussing 
the behavior in the critical region at nonzero noise.  This is available
if one includes the gaussian fluctuations in the saddle point 
calculation of eqs.(\ref{mus}). 
If one defines $g_c(D)=\frac{3}{\sqrt{2}}\left(1+\frac{D}{2}\ln\frac{D}{2}
\right)$ and $\epsilon \equiv g-g_c$, the result can be
cast in the scaling form 
$s_{\infty}=D f(\epsilon/D)$,
for $\epsilon$ and $D$ both small, where $f(x)$ satisfies 
$3(1+f+\ln\sqrt{f-1})=\sqrt{2}x$.
A similar scaling form holds for $\mu_{\infty}=-\sqrt{2}D-s_{\infty}/\sqrt{2}$.
For $D>0$ the transition is smoothed and a crossover between the two
regimes occurs. We checked by a numerical integration of the Langevin 
eq.\ (\ref{modorl}) for $\sim 10^6$ realizations (see fig. \ref{fig1})
that the stationary state discussed so far is the one which is actually 
reached by the dynamics.

Let us discuss our findings in the framework of kinetic roughening.
The parameter $s$ has the meaning of the square of the interface width
$s=\langle{(h-\langle{h}\rangle)^2}\rangle$. 
Its behavior with the linear system size
$L$, in the stationary state, defines the roughness exponent $\chi$:
$s_{\infty}(L)\sim L^{2\chi}$. In our case, we considered from the beginning an
infinite system, so a finite value of $s_{\infty}$, both in the weak
and in the strong coupling phases, implies $\chi=0$. 
This agrees with the result found for  
directed polymers in random media in the $d\to\infty$ limit\cite{spohn}. 
However, the nature of the stationary state differs dramatically.
Indeed eq.\ (\ref{P(phi)}) predicts, for the strong coupling regime, a
double peaked distribution in contrast to the highly asymmetric but
single peak distribution found for directed polymers\cite{dprmfsc}. The
most striking feature of the distribution (\ref{P(phi)}) is that it
predicts the presence of bumps of a definite height $\phi^+-\phi^-$ in
the surface for $g>g_c$. Moreover, the minima of the potential are 
separated by a barrier 
$\Delta V\simeq(g/3)^4$ which is very 
large for $g \gg 1$. This suggests very large relaxation
times $\tau\sim \exp(\Delta V/D)$ in the stationary state. The
numerical simulation of eq.\ (\ref{modorl}) indeed confirms this
conclusion (see fig. \ref{fig1}). The interface velocity, 
$v_\infty=gs_\infty - \mu_\infty$, also differs in the two regimes, 
being $O(D)$ for $g<g_c$ and $O(g-g_c)$ for $g>g_c$.

Let us now discuss the non-stationary dynamics in the strong-coupling 
regime. Consider the Langevin dynamics of an ensemble of 
a $z\gg 1$ system $\{\phi_i\}$ starting from a flat substrate: 
$P_{t=0}(\phi)=\delta(\phi)$. 
Therefore $s_0=0=\mu_0$. It is convenient to think again in terms of the 
potential (\ref{V(phi)}). Initially this has two minima, but
only the minimum at $\phi^-_0=0$ is occupied, while the one at 
$\phi^+_0\simeq g$, which is very deep, with $V_0(\phi^+)\simeq -g^4/12$, 
is empty. 
As time goes on, the locations of the minima, which are given by 
$V_t'(\phi_t^\pm)=0$, adjust according to the running values
of $s_t$ and $\mu_t$. At the same time, the potential barrier 
between the two minima increases. 
After an initial transient, the distribution of the ensemble 
$\{\phi_i\}$ can be well approximated by 
$P_t(\phi)=\alpha_t\delta(\phi-\phi^+_t)+(1-\alpha_t)
\delta(\phi-\phi^-_t)$. This allows one to transform the equations
$\langle{\phi}\rangle_t=0$ and $s_t=\langle{\phi^2}\rangle_t$ in the form 
$\alpha_t=-\phi^-_t/(\phi^+_t-\phi^-_t)$ and $s_t=-\phi^-_t\phi^+_t$.
These relations leave only one independent variable,  
whose dynamics is given by
\begin{equation}
\frac{d\alpha}{dt} \simeq \exp\left[-
\frac{V_t(\phi^o_t)-V_t(\phi^-_t)}{D}\right].
\label{dynamics}
\end{equation}
This equation describes the noise activated processes $\phi^-\to\phi^+$
across the potential barrier. It neglects the
backward processes $\phi^+\to\phi^-$. It is therefore valid as long as
$V_t(\phi^-)-V_t(\phi^+)\gg D$. Since the scale of $V_t$ is of order 
$g^4\gg D$, this is a good approximation for times for which
the system is not too close to the stationary state.

The equation (\ref{dynamics}) cannot be solved exactly. 
However some progress can be made assuming $\alpha\ll 1$, 
and considering the leading behavior in $\alpha$: 
$\phi^+=g(1-\alpha)$, $\phi^-\simeq -g\alpha$ and
$\phi^o\simeq 2g\alpha$. To leading order in $\alpha\ll 1$, after some 
algebra, one finds
\[\frac{V_t(\phi^o_t)-V_t(\phi^-_t)}{D}=
\frac{9g^4\alpha^3}{2D}=\left(\frac{\alpha}{t_c}\right)^3\]
where the characteristic time
$t_c=\left(\frac{2D}{9g^4}\right)^{1/3}$ has been defined.
Note that in the stationary state $\alpha_\infty\to 0.1464\ldots$
as $g\to\infty$. Therefore the leading order in $\alpha_t$
can give a reasonable approximation even for late times.

For $\alpha_t\ll t_c\ll 1$, the saddle point approximation 
to $P_t(\phi)$ becomes questionable. 
In the regime $t_c \ll \alpha_t\ll 1$, where we expect our
approximations to be reasonable, eq.\ (\ref{dynamics}), to
leading order in $\alpha_t$, predicts a behavior
$\alpha_t\sim (\ln t)^{1/3}$.
The ``interface width'' $s_t$ can easily be 
related to $\alpha_t$. Within our approximations
($g^4\gg D$ and $\alpha_t\ll 1$) one finds
$s_t\simeq D+(\phi^+)^2\alpha_t$
to leading order in $\alpha_t$.
Therefore our previous discussion suggests 
\begin{equation}
s_t\sim (\ln t)^{1/3}.
\label{alog}
\end{equation}
Figure \ref{fig1} shows that for $g=3$ and $D=0.0625$ eq.\ (\ref{alog}) 
yields an accurate fit of the simulation results. 

This regime holds in general for $D\ll g^4$ (when the saddle point
approximation is valid) and for intermediate times. 
When the backward processes
$\phi^+\to\phi^-$ becomes important ($V(\phi^-)-V(\phi^+)\sim D$), 
eq.\ (\ref{dynamics}) ceases to hold. The system starts to relax to
the stationary state (see figure \ref{fig1}).

It is important to note that the `off equilibrium' dynamics has
no relation whatsoever with the stationary dynamics. In particular 
we found that the relaxation times {\em to} the stationary state 
are orders of magnitude smaller than the ones which characterizes 
the dynamics {\em in} the
stationary state. In the case illustrated in figure \ref{fig1},
for example, the fit of the approach to the stationary state yields a 
characteristic time $\approx 340$, whereas the correlation function in 
the stationary state decays over characteristic times of order 
$\exp(\Delta V/D) \approx 8000$.
The scaling relation usually assumed to relate the early stages of
growth to the stationary state behavior\cite{kpzrev} fails
dramatically. 
Finally we observed that the time scales involved in the dynamics of the
model, even for moderate values of $g$ and $D$, can well be beyond 
the reach of numerical simulation (already for $g=4$ and $D=0.125$ the
time to reach the stationary state is of order $10^5$). 

In conclusion, we have discussed a high-dimensional limit for growth
equations. We have shown that this limit is plagued by finite time
singularities for the KPZ equation as it stands. We have modified
the model by including a further term in the small gradient expansion of
the equation for RSOS growth models. Even though this term would be
irrelevant, according to the usual folklore based on dimensional
analysis, we have shown that it allows for a well defined and non-trivial 
high dimensional limit. In particular, in the stationary state
the resulting surface is characterized by bumps (or depressions, for 
$g < -g_c$ \cite{note}) of a definite height in the strong-coupling regime. 
Although the analysis has been carried out explicitly in the limit of 
vanishing noise, $D=0^+$, the same qualitative picture obtains for $D>0$, 
with the height distribution changing (but now smoothly) from single-peaked 
to double-peaked with increasing $g$. 

This strong-coupling stationary state differs substantially from the 
one of directed polymers in random media. 
Our results suggest that the
mapping discussed in ref.\ \cite{mapping} between RSOS models and
directed polymers may fail in high dimensions. 
One might wonder whether the scenario outlined so far is specific to 
eq. (\ref{kpzRSOS}) or whether it applies in a more general class of
models. Note, in this respect, that the sign
of $\kappa$ plays a very important role. For $\kappa<0$ the 
infinite dimensional limit is plagued by finite time singularities
like those found for the KPZ equation. Actually, eq. (\ref{kpzRSOS}) with 
$\kappa<0$ would result from the small gradient expansion of the
equation for the Eden model or for ballistic aggregation \cite{mmtb}.
On the other hand, for other generalizations of the KPZ
equation  (like e.g. the one used in ref.\cite{esipov}) in the spirit of
RSOS models, we found the same scenario as in this work \cite{boundedV}.  
These observations raise the question of whether a KPZ universality class
exists at all in high enough dimensions. 
It is interesting to note that already for 
$d=2$ there are discrepancies between the exponents obtained
by numerical simulations for different models \cite{num2}. 
It is also noteworthy that already in $d=2$ finite-time singularities in 
the simulation of the KPZ equation have been observed\cite{Kim}. 

If the behavior discussed above applies to any
large but finite dimension, a second key question is what is the
dimension $d_c$ above which this behavior sets in. In this respect we 
note that finite height bumps and structures with very large relaxation 
times have been observed\cite{Tim} in $d=2$ simulations of a 
{\it regularized} KPZ equation. 

We finally point out that the existing numerical results
on RSOS models \cite{ala} in high dimensions do not completely rule
out the possibility that the behavior we have discussed sets in for
relatively low dimensions. Indeed in ref.\ \cite{ala} only the
early stages of growth were investigated and an exponent was extracted 
{\it under the assumption of dynamic scaling}. Our model shows
that this assumption is far from trivial. Furthermore, in a numerical 
simulation for the interface thickness it might be quite hard to 
disentangle a power law with a small power from a behavior like that 
of eq.\ (\ref{alog}).

We thank T. Blum, M. A. Moore, and T. J. Newman for discussions.  
This work was supported by EPSRC grant GR/H27496.

\begin{figure} 
\caption{Upper curve: Stationary state correlation function 
$C(t)\equiv\langle{\phi(t+t_0)\phi(t_0)}\rangle$
for a system with $g=3$ and $D=1/16$.
The fit to $C(t)=A\exp(-t/\tau)$ yields $\ln \tau\simeq 8.97$ to be compared 
to $\Delta V/D = 9$.
Lower curve: behavior of $s_t$, starting from the flat
configuration, for the same system. The fit was obtained 
plotting $s^3\simeq 0.015 + 0.017 \ln t$ versus $\ln t$.}
\label{fig1}
\end{figure}


\begin{references}
\bibitem{note} For RSOS models, $\lambda <0$. The invariance of 
(\ref{kpzRSOS}) under $\lambda \to -\lambda$, $h \to -h$, implies 
however that the sign of $\lambda$ is unimportant, and we take 
$\lambda > 0$ throughout.
\bibitem{nota} The $\lambda$ term is a correction to a constant $f$ 
    term, which can be eliminated by a Galilean transformation.
\bibitem{boundedV} The key stability requirement is that the potential 
$V(\phi)$ be bounded below.
\end{references}
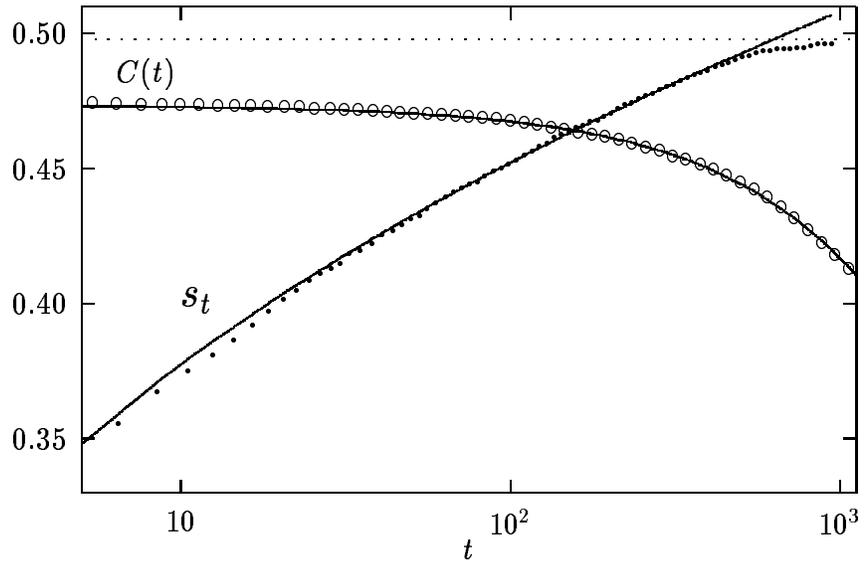
\end{document}